# SNKnock: A free security tool for Facebook users

Mahmoud Samir Fayed
King Saud University
Riyadh, Saudi Arabia
mafayed@KSU.EDU.SA

*Abstract*—The Facebook Social network is very famous and widely used by millions of users all over the world. Facebook comes with high level of usability so users can easily find their friends and connect to them, but there are security issues related to this process where the attacker can make same-site or cross-site profile cloning attacks to get other users data. In this paper we will identify advanced same-site or cross-site profile cloning attacks then we will propose a security solution to these attacks where the user must record a voice message and answer some questions when he sends a friend request to another user, and then the other user will listen to this message before deciding to accept or reject the request. We implemented this solution by developing new web application called SNKnock which is available as free service.

## I. INTRODUCTION

The social network becomes a famous and well known tool for communication with the peoples which we know in the real world and, The Facebook website is an example of a very successful social network (SN) with millions of users all over the world [1].

During the last years the Facebook social network website becomes a general direction for research by many researchers all over the world. And now there are many studies in different dimensions around the Facebook, for example there are studies in the human behavior [2-9], Psychology [10], business [15-16], decision support systems [17] and security [11-14].

To start using facebook you need to create an account with a username and password, then you start by adding your information (optional) like data of birth, job & photos then you search for your friends by their names or emails [20].

To communicate with your friends you can add them to your friends list and this process requires their permission where you send the friend request and then your friend accepts that request and after that only you can see their posts, photos & links which are set by them to be visible only to the accounts in their friends list [18] Before being a friend you can only see public posts and nothing more. Public posts may include their posts in the group you are joined and/or their comments on the mutual friend's posts [19].

The problem with this procedure (Send friend Request - then Accept/Reject) are concentrated in the same-site and/or cross-site profile cloning attacks, where the attacker can create an account and put the name of a person known to the victim then send the friend request which is expected to be accepted by the victim because he/she think it's coming from a person they know then the attacker gets his/her information [11].

If the account created by the attacker is the first account with this name in the same social network , i.e. the real user don't have account in this social network, then it's cross-site profile cloning attack which don't lead to suspicion. But, if the account is duplication to a real account in the same social network then it's a same-site profile cloning attack which may lead to suspicion and discovered easily.

In this article we will concentrate on the profile cloning attacks and we will identify advanced types of these attacks then we will propose a security solution to this attack.

The structure of this paper is

**Section2**: Profile cloning attacks and users awareness
**Section3**: Identification of advanced profile cloning attacks
**Section4**: The proposed solution
**Section5**: The implementation
**Section6**: Conclusion and Future work

## II. PROFILE CLONING ATTACKS AND USERS AWARENESS

When the user know about profile cloning attacks he/she can do a check before accepting the friend request, The check done by opening the profile page of the person who sent the friend request and looking at the data to guess is this a real account or not, This could includes for example

- Looking at other information other than the name and the photo, this information maybe information like data of birth, job, city & more photos.
- Looking at the public posts/comments by this user and when he/she joined the Facebook

- Looking at comments of other users on the public posts
- Looking at the mutual friends

The attacker or the user may hide most of this information and make it visible only to friends. The reason if this is an attacker is to do a trick while if this is a real user he/she may do this for security and privacy reasons. So when the person don't see enough information he can't quickly guess is this a real account or not, and may accept the friend request then do another check to look at more information.

After accepting the friend request the vision becomes clearer and the information may be enough to take the decision to keep this user as a friend or to delete it.

Some users may look only at the available information and concentrate on the important ones like the mutual friends and take final decision based on that.

### III. IDENTIFICATION OF ADVANCED PROFILE CLONING ATTACKS

Now we will define the advanced profile cloning attack which takes advantage from the next behaviors

1 – Some users may decide to accept or reject the friend request based only on the available information and trust the account based on the mutual friends
2 – The attacker can change the name which appears in the profile
3 – The attack can be done in more than one stage
In the advanced profile cloning attack, the attacker could do the attack in the next stages.

*The First Stage (Prepare)*

- Create a group of Facebook accounts (using more than one account in the attack is an idea from the Sybil attack [11] which get advantage of the structure of the social network) this group will be called LIST(1). Look at figure (1).
- Connect these accounts together in a one to many structures where one account will be the head or the center and will be called Attacker (Root) which contains profile page with a lot of posts and comments from other fake accounts.
- Spend some time making a communication between these fake accounts until the Attacker(Root) account seems to be an account for real human have friends and is doing real activities

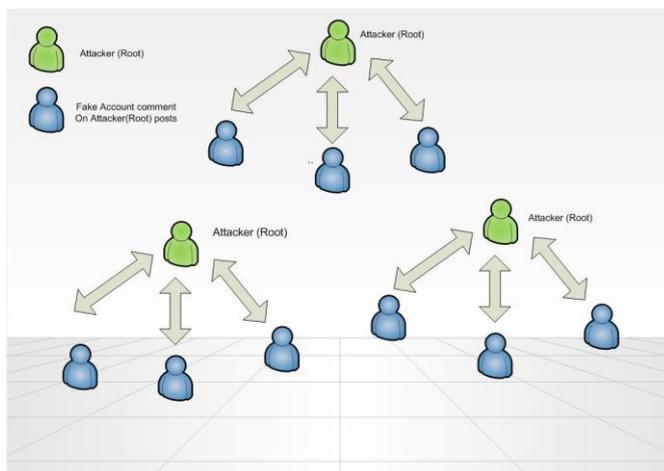

Fig. 1 The first stage in the advanced profile cloning attack

*The Second Stage (Hidden movement)*

- Know as much as you can of accounts which is connected to the victim from the friend list (if it's visible) or by doing a search using names (if the accounts appears in the search results) and this list will be called LIST(2). Look at figure (2)

- Use some of the fake accounts and send friend requests to the collected accounts and see if they will accept the friend request or not
    - If Yes
        - Record this account as a weak account, send the friend request directly from all of the fake accounts in List(1) including the Attacker(Root) account to become a friend
    - If you can't find a weak accounts do the previous procedure with every account in List(2)

*The third stage (Grow)*

- Repeat stage one and two for more than one time to prepare more than one network Attacker(Root) and List(1) to use in the next stage
- Use your experience from the previous stages to quickly create these accounts and connect it to the accounts in List(2)

*The Last stage (Go to the victim)*

- Rename the attacker(Root) accounts to a names known by the victim ( use 2 or 3 attaker(Root) )
- Send the friend requests at the same time or one by one (try both if you have many attacker(Root) accounts)

Now the victim will find these accounts carry names known by him/her and there is activity by these accounts and there are real mutual friends. One of these requests could be accepted based on the account name and the relationship between the victim and the real person whose name is used.

The goal from the identification of this attack is to clear that even if the user is aware of the profile cloning attack it's not enough to check the account before and after accepting the friend request because the attacker can put data and can make mutual friends to let us figure that everything is ok

We need a more secure system when we accept the friend requests.

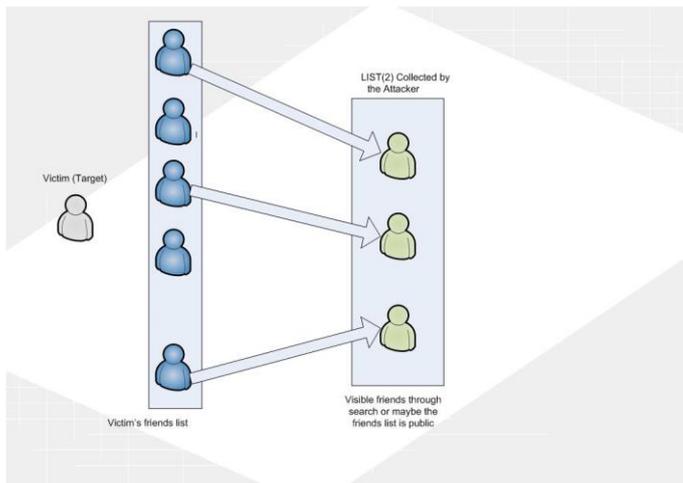

Fig. 2 List(2) collected by the attacker

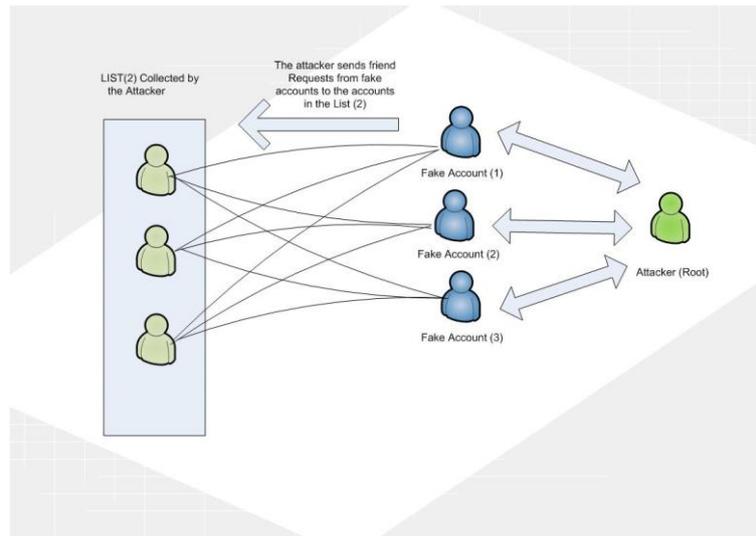

Fig. 3 The attacker is trying at first to find the weak accounts in the List(2)

IV. THE PROPOSED SOKUTION

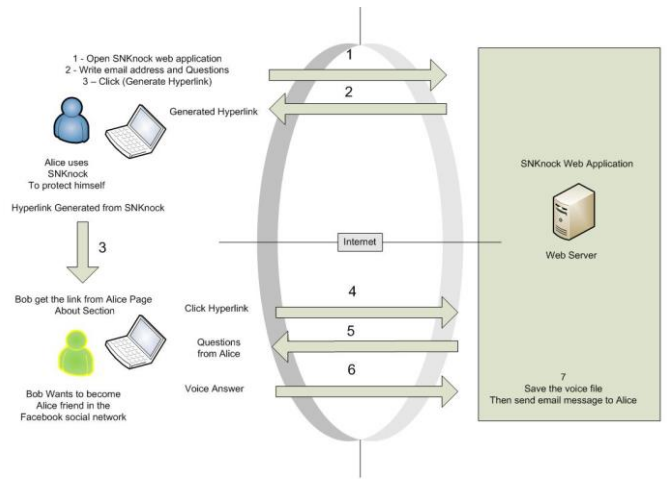

Fig.4 SNKnock Model

We propose a security solution to the profile cloning attacks. Our proposal is based on character identification through their voice (Something you are).

We will develop a web application and the user who wants to protect hiself/herself against the profile cloning attacks will open this web application. The web application will ask the user to determine 2 things as input (email address & questions) then will generate a link which will be the output of the web application.

After that the user will spread this hyperlink to peoples looking to become his/her friends in the social network. When someone would like to become a friend he/she will click on that link then will answer the questions using his/her voice.

After that an email message will sent to the original user contains a link to the sound file. The original user will listen to the sound file then decide to accept or reject the friend request.

### VE. THE IMPLEMENTATION (THE SNKNOCK WEB APPLICATION)

We implemented our proposed solution by developing a web application called SNKnock[21] which developed using HTML [22], CSS [23], JavaScript [24], PHP[25] & MySQL[26].

The SNKnock user interface is available in two languages (Arabic & English) and the user can change the language using a hyperlink in the bottom of the webpage.

The SNKnock database contains 1 table called Questions
The table contains the next columns

TABLE1. THE STRUCTURE OF THE QUESTION TABLE

| Index | Column Name | Type |
|---|---|---|
| 1 | Id | Int (AUTO_INCREMENT) |
| 2 | Email | Varchar |
| 3 | questions | Varchar |

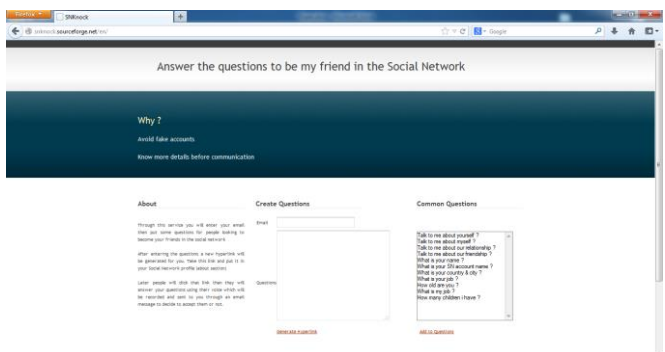

Fig.5 SNKnock web application

In the front page of SNKnock the user type his/her email address then enter one or more of questions.

There is a Listbox contains questions suggestions to be used by the user like

- Talk to me about yourself?
- Talk to me about myself?
- Talk to me about our relationship?
- Talk to me about our friendship?
- What is your name?
- What is your SN Account Name?
- What is your country & city?
- What is your job?
- How old are you?
- What is my job?
- How many children I have?

The user can mix between typing questions and selecting questions from the listbox. The next figure present a real example demonstrates how to use the service

The user entered an email address: Alice@test.com Then selected/typed one question: Talk to me about myself?

After that the user clicked on the hyperlink: Generate Hyperlink. As a result for this a new hyperlink is generated: http://snknock.sf.net/en/answer.php?code=43 . The hyperlink generation process done by the next php source code

```
<HEAD>

<?PHP
$email= $_POST['user_email'];
$questions= $_POST['user_questions'];
?>

</HEAD>

<?php

 include('snknock_db.php');
 $con = snknock_db_open();
if (mysqli_connect_errno())
{
    echo "Failed to connect to MySQL: " .
    mysqli_connect_error();
 }
mysqli_query($con,"INSERT INTO questions
(id , email, questions) VALUES ('' , '" . $email .
"', '" . $questions . "')");
$nCode = mysqli_insert_id($con) ;
echo "<p style='text-align:center'>" ;
echo "Generated link : <a
href='http://snknock.sf.net/en/answer.php?code="
. $nCode . "' target='_blank'>
http://snknock.sf.net/en/answer.php?code=" .
$nCode . "</a> ";
echo "<br />" ;
echo "</p>" ;
mysqli_close($con);

?>
```

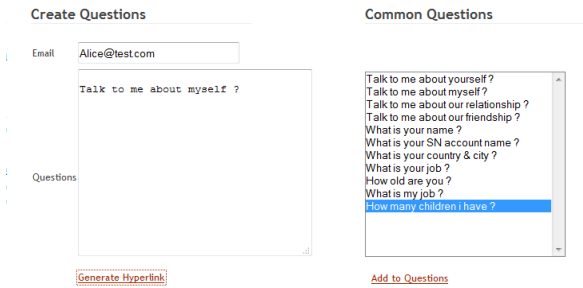

Fig. 6 Generate new hyperlink using SNKnock

The user now can put this hyper link in his/her profile page in the Facebook website. The link can be inserted to the about section along with a message like (Please, to be my friend, send the friend request then click on this link and answer my question using your voice, I do that because I would like to avoid fake accounts and profile cloning attacks) or any other message by the user.

When another user or an attacker click on the link he will get a webpage similar to the one in the next figure

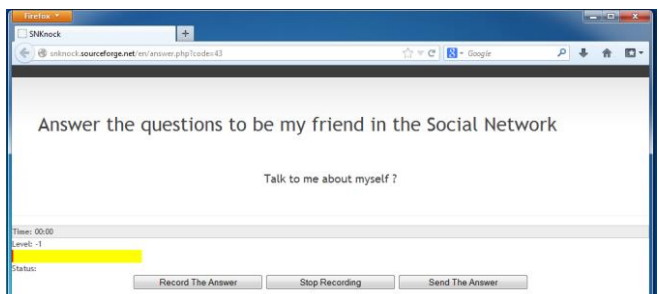

Fig. 7 recording the answer using SNKnock

The user or the attacker will see the question and he will answer it using his/her voice. To answer the question he/she will click the button "Record the answer" then will answer the question then will click "Stop recording" and he will listen to his/her answer.

When the user or the attacker click "send the answer" the recorded voice will be saved in a sound file stored on the web server and an email message will be sent to the original user who generated the hyperlink to this webpage using SNKnock.

The questions are displayed by the next PHP code

```
<HEAD>

<?PHP
$code = $_GET["code"];
?>

</HEAD>

<?php

 $str_file = file_get_contents( "header.html" ) ;
 echo $str_file ;
 include('snknock_db.php');
 $con = snknock_db_open();
 if (mysqli_connect_errno())
 {
     echo "Failed to connect to MySQL: " .
     mysqli_connect_error();
 }
 $result = mysqli_query($con,"SELECT * FROM questions
 where id = '" . $code . "'");
 while($row = mysqli_fetch_array($result))
 {
        echo " <p style='text-align:center'> <font size='4'> " ;
        echo "<br />";
        foreach(preg_split("/((\r?\n)|(\r\n?))/",
        $row['questions']) as $line)
        {
               echo $line;
               echo "<br />";
        }
        echo " </font> </p > " ;
        $cont = "1" ;
}
if ( $cont == "1" )
{
    $str_file = file_get_contents( "voice.html" ) ;
    $str_file = str_replace("THEFILENAMEISHERE",
    "answerfile_" . rand(1,2000) .
    rand(1000,2000) . rand(2000,5000) . rand(7000,90000) ,
    $str_file);
    $str_file = str_replace("THEUSERCODEISHERE",
    $code , $str_file);
    echo $str_file ;
}

mysqli_close($con);

?>
```

After that the original user will get an email message, he/she will click on the sound file link which is embedded in the email message to listen to the answer and after that he will decide to accept or reject the friend request.

CONCLUSION AND FUTURE WORK

In this paper we identified advanced profile cloning attacks and proposed a security solution to this type of attack then we implemented our proposed solution by developing a web application.

We know that there are some limitations to our solution for example some users don't have mice device to record the sound, some hackers may find a way to pass this voice test, sometimes the Facebook user get friend requests from people known to him but he/she never listened to their voice before.

In the future we are looking to extend our solution by fixing known issues and adding more security options like identification based on web camera and nested questions where the user will get a question based on his answer to the previous question.